\documentclass[%
	final,
	journal,%conference,% conference, , technote, peerreview, peerreviewca
    comsoc,%comsoc, compsoc, transmag
	letterpaper,% letterpaper, a4paper, cspaper
	oneside,% oneside, twoside
	twocolumn,% onecolumn, twocolumn % IEEE uses always twocolumn % onecolumn just for info.
	nofonttune,%
]{IEEEtran}%
\PassOptionsToPackage{usenames,dvipsnames,svgnames,x11names,table,prologue}{xcolor}%
\PassOptionsToPackage{hyphens}{url}%
\PassOptionsToPackage{nomessages}{fp}%
\ifCLASSINFOpdf
  \usepackage[pdftex]{graphicx}
\else
  \usepackage[dvips]{graphicx}
\fi
\ifCLASSOPTIONcompsoc
   \usepackage[caption=false,font=normalsize,labelfont=sf,textfont=sf]{subfig}
\else
   \usepackage[caption=false,font=footnotesize]{subfig}
\fi

\usepackage{stfloats}
\usepackage[english]{babel}%
\usepackage[utf8]{inputenc}%
\usepackage[babel,style=english]{csquotes}%
\usepackage{hyphsubst}%
\usepackage[%
	activate={true,%
	nocompatibility},%
	final,%
	tracking=true,%
	kerning=true,%
	spacing=true,%
	factor=1100,%
	stretch=10,%
	shrink=10%
]{microtype}%
\usepackage{setspace}%
\usepackage{xcolor}%
\definecolor{todonotecol}{RGB}{250,0,0}%
\usepackage{xparse}
\usepackage[%
	colorlinks=false,%
	urlcolor=black,%
	linkcolor=black,%
	citecolor=black,%
	filecolor=black,%
	breaklinks,%
	]{hyperref}%
\usepackage{url}%
\usepackage[%
	acronym,%
	nopostdot,%
	seeautonumberlist,%
	shortcuts,%
	section=chapter,%
	toc,%
]{glossaries}%
\glsaddkey*{url}%Key Name
{}%Default Value
{\glsentryurl}%Use analog \glsentrytext
{\Glsentryurl}%Use analog \Glsentrytext
{\glsurl}%Use analog \glstext
{\Glsurl}%Use analog \Glstext
{\GLSurl}%Use analog \GLStext%
\glsaddkey*{longpltwo}%Key Name
{}%Default Value
{\glsentrylongpltwo}%Use analog \glsentrytext
{\Glsentrylongpltwo}%Use analog \Glsentrytext
{\glslongpltwo}%Use analog \glstext
{\Glslongpltwo}%Use analog \Glstext
{\GLSlongpltwo}%Use analog \GLStext%
\newglossaryentry{}{%
	name={},%
	plural={},%
	description={},%
}%
\newacronym{it}{IT}{information technology}
\newacronym{ot}{OT}{operational technology}
\newacronym{opcua}{OPC UA}{Open Platform Communications Unified Architecture}
\newacronym{tsn}{TSN}{IEEE time-sensitive networking}
\newacronym{scada}{SCADA}{supervisory control and data acquisition}
\newacronym{vpc}{VPC}{Virtual Process Controller}
\newacronym{plc}{PLC}{Programmable Logic Controller}
\newacronym{udp}{UDP}{User Datagram Protocol}
\newacronym{pubsub}{PubSub}{Publish Subscribe}
\newacronym{vm}{VM}{virtual machine}
\newacronym{pc}{PC}{personal computer}
\newacronym{ptp}{PTP}{Precision Time Protocol}
\newacronym{gptp}{gPTP}{generalized Precision Time Protocol}
\newacronym{ntp}{NTP}{Network Time Protocol}
\newacronym{splc}{Soft-PLC}{Software-PLC}
\newacronym{rte}{RTE}{Real-Time Ethernet}
\newacronym{osi}{OSI}{Open Systems Interconnection}
\newacronym{mac}{MAC}{Media Access Control}
\newacronym{e2e}{E2E}{end-to-end}
\newacronym{tcp}{TCP}{Transmission Control Protocol}
\newacronym{hmi}{HMI}{Human-Machine Interface}
\newacronym{kpi}{KPI}{Key Performance Indicator}
\newacronym{mes}{MES}{Manufacturing Execution System}
\newacronym{rami4.0}{RAMI 4.0}{Reference Architectural Model Industrie 4.0}
\newacronym{tacnet4.0}{TACNET 4.0}{Tactile Internet 4.0}
\newacronym{3gpp}{3GPP}{3rd Generation Partnership Project}
\newacronym{5gppp}{5GPPP}{5G Infrastructure Public Private Partnership}
\newacronym{itu}{ITU}{International Telecommunication Union}
\newacronym{ngnm}{NGNM}{Next Generation Mobile Networks}
\newacronym{agv}{AGV}{automated guided vehicle}
\newacronym{v2v}{V2V}{vehicle-to-vehicle}
\newacronym{v2x}{V2X}{vehicle-to-infrastructure}
\newacronym{d2x}{D2X}{device-to-x}
\newacronym{qos}{QoS}{quality of service}
\newacronym{noa}{NOA}{NAMUR Open Architecture}
\newacronym{dcs}{DCS}{distributed control system}
\newacronym{m2m}{M2M}{machine-to-machine}
\newacronym{sdn}{SDN}{software-defined networking}
\newacronym{erp}{ERP}{enterprise resource planning}
\newacronym{ip}{IP}{Internet Protocol}
\newacronym{lte}{LTE}{Long Term Evolution}
\newacronym{5g}{5G}{5th generation wireless communication systems}
\newacronym{harq}{HARQ}{Hybrid Automatic Repeat Request}
\newacronym{rat}{RAT}{radio access technology}
\newacronym{mc}{MC}{Multi-Connectivity}
\newacronym{fbmc}{FBMC}{Filter Bank Multicarrier}
\newacronym{gfdm}{GFDM}{Generalized Frequency Division Multiplexing}
\newacronym{sla}{SLA}{service-level agreement}
\newacronym{5qi}{5QI}{5G QoS Indicator}
\newacronym{bmbf}{BMBF}{German Federal Ministry of Education and Research}
\newacronym{nrt}{NRT}{non real-time}
\newacronym{irt}{IRT}{isochronous real-time}
\newacronym{rt}{RT}{real-time}
\newacronym{ar}{AR}{augmented reality}
\newacronym{wlan}{WLAN}{wireless local area network}
\newacronym{cpu}{CPU}{central processing unit}
\newacronym{ram}{RAM}{random-access memory}
\newacronym{ue}{UE}{user equipment}
\newacronym{ran}{RAN}{radio access network}
\newacronym{bs}{BS}{base station}
\newacronym{cn}{CN}{core network}
\newacronym{epc}{EPC}{evolved packet core}
\newacronym{mec}{MEC}{mobile edge cloud}
\newacronym{rtt}{RTT}{round-trip time}
\newacronym{vpn}{VPN}{virtual private network}
\newacronym{vpf}{VPF}{Virtual Process Control Function}
\newacronym{vdi}{VDI}{Association of German Engineers}
\newacronym{ipc}{IPC}{industrial pc}
\newacronym{rtos}{RTOS}{real-time operating system}
\newacronym{mano}{VPC M\&O}{VPC Manager \& Orchestrator}
\newacronym{i/o}{I/O}{Input / Output}
\newacronym{fbd}{FBD}{Function Block Diagram}
\newacronym{ldd}{LDD}{Ladder Logic Diagram}
\newacronym{sfc}{SFC}{Sequential Function Chart}
\newacronym{il}{IL}{Instruction List}
\newacronym{st}{ST}{Structured Text}
\newacronym{os}{OS}{Operating System}
\newacronym{mqtt}{MQTT}{message queuing telemetry transport }
\newacronym{amqp}{AMQP}{advanced message queuing protocol}
\newacronym{api}{API}{application programming interface}
\newacronym{embb}{eMBB}{enhanced mobile broadband}
\newacronym{iot}{IoT}{Internet of Things}
\newacronym{urllc}{URLLC}{ultra-reliable low-latency communication}
\newacronym{mmtc}{mMTC}{massive machine-type communications}
\newacronym{iiot}{IIoT}{industrial IoT}
\newacronym{iiot2}{IIoT}{industrial Internet of Things}
\newacronym{npn}{NPN}{non-public network}
\newacronym{prb}{PRB}{Physical Ressource Block}
\newacronym{gnb}{gNB}{5G base station}%next generation Node B}
\newacronym{5gs}{5GS}{5G system}
\newacronym{5gc}{5GC}{5G core network}
\newacronym{oai}{OAI}{OpenAirInterface}
\newacronym{enb}{eNB}{4G base station}
\newacronym{sdr}{SDR}{Software Defined Radio}
\newacronym{usrp}{USRP}{Universal Software Radio Peripheral}
\newacronym{mme}{MME}{Mobility Management Entity}
\newacronym{hss}{HSS}{Home Subscriber Server}
\newacronym{pdn}{PDN}{Packet Data Network}
\newacronym{sgw}{SGW}{Serving Gateway}
\newacronym{gtp}{GTP}{GPRS Tunneling Protocol}
\newacronym{mimo}{MIMO}{Multiple Input Multiple Output}
\newacronym{fpga}{FPGA}{Field Programmable Gate Array}
\newacronym{pss}{PSS}{primary synchronization signal}
\newacronym{sss}{SSS}{secondary synchronization signal}
\newacronym{sfn}{SFN}{system frame number}
\newacronym{mib}{MIB}{master information block}
\newacronym{pbch}{PBCH}{Physical Broadcast Channel}
\newacronym{ism}{ISM}{Industrial, Scientific and Medical}
\newacronym{e-utra}{E-UTRA}{evolved UMTS Terrestrial Radio Access}
\newacronym{bldc}{BLDC}{brushless DC}
\newacronym{dl}{DL}{Downlink}
\newacronym{ds-tt}{DS-TT}{Device-Side TSN Translator}
\newacronym{nw-tt}{NW-TT}{Network-Side TSN Translator}
\newacronym{upf}{UPF}{User Plane Function}
\newacronym{tsi}{TSi}{ingress timestamping}
\newacronym{tse}{TSe}{egress timestamping}
\newacronym{gm}{GM}{grandmaster}
\newacronym{sn}{SN}{sequence number}
\newacronym{rbis}{RBIS}{Reference Broadcast Infrastructure Synchronization}
\newacronym{erbis}{ERBIS}{Extended Reference Broadcast Infrastructure Synchronization}
\newacronym{cococo}{CoCoCo}{communication-control codesign}
\newacronym{3c}{3C}{joint design of communications, control, and computing}
\newacronym{cots}{COTS}{commercial off-the-shelf}
\newacronym{ie}{IE}{industrial Ethernet}
\newacronym{ap}{AP}{access point}
\newacronym{ofdm}{OFDM}{Orthogonal Frequency-Division Multiplexing}
\newacronym{csma}{CSMA-CA}{carrier sense multiple access with collision avoidance}
\newacronym{dcf}{DCF}{distributed coordination function}
\newacronym{pcf}{PCF}{point coordination function}
\newacronym{hcf}{HCF}{hybrid coordination function}
\newacronym{edca}{EDCA}{enhanced distributed channel access}
\newacronym{hcca}{HCCA}{\gls{hcf} controlled channel access}
\newacronym{tdma}{TDMA}{time division multiple access}
\newacronym{ofdma}{OFDMA}{Orthogonal Frequency-Division Multiple Access}
\newacronym{mumimo}{MU-MIMO}{multi-user \gls{mimo}}
\newacronym{bi}{BI}{beacon interval}
\newacronym{ssid}{SSID}{service set identifier}
\newacronym{icps}{ICPS}{industrial cyber-physical system}
\glsdisablehyper% Disables Hyperlinks
\usepackage[%
	backend=biber,%
	style=ieee,%
	isbn=false,%
	hyperref=true,%
	maxbibnames=99,%
	sorting=none,%
	natbib=true,%
	language=english,%
	defernumbers=true,%
	]{biblatex}%
\DeclareFieldFormat{sentencecase}{\csname bbx@colon@search\endcsname#1}% Therewith capital letters stay capitals, when using 'style=ieee'

\usepackage{nameref}
\usepackage{soul}
\usepackage{flushend}
\usepackage{textcomp}
\usepackage{calc}
\usepackage{xkeyval}
\usepackage{multirow}
\usepackage{tabulary}% Nearly the same as 'tabularx', bit more extensive
\usepackage{makecell}% http://ctan.org/pkg/makecell
\usepackage{pgfplots}
\usepackage{tikz}
\usepackage{textcomp} 
\usepackage{verbatim}
\newcommand{\nl}{\par\noindent} % Abk. für NewLine
\newcommand{\mytilde}{{\raise.17ex\hbox{$\scriptstyle\mathtt{\sim}$}}}

\newlength\textheighttemp%
\newlength\textwidthtemp%
\newlength\textheightstd%
\newlength\textwidthstd%
\newlength\textheightold%
\newlength\textwidthold%
\newlength\tempheight%
\newlength\tempwidth%
\SetProtrusion{encoding={*},family={bch},series={*},size={6,7}}
              {1={ ,750},2={ ,500},3={ ,500},4={ ,500},5={ ,500},
               6={ ,500},7={ ,600},8={ ,500},9={ ,500},0={ ,500}}
\SetExtraKerning[unit=space]
    {encoding={*}, family={bch}, series={*}, size={footnotesize,small,normalsize}}
    {\textendash={400,400}, % en-dash, add more space around it
     "28={ ,150}, % left bracket, add space from right
     "29={150, }, % right bracket, add space from left
     \textquotedblleft={ ,150}, % left quotation mark, space from right
     \textquotedblright={150, }} % right quotation mark, space from left
\SetTracking{encoding={*}, shape=sc}{40}
\makeatletter
\let\blx@rerun@biber\relax
\makeatother

\usepackage{newtxmath}
\pgfplotsset{
  grid style = {
   line width = 0.1pt
  }
}
				\newcommand{\disablewr}[1]{#1}%
				\newcommand{\newcommanddisw}[3]{\newcommand{#1}[1]{\disablewr{\textcolor{#2}{#3}}}}%
\renewcommand{\disablewr}[1]{}%
\definecolor{todocol}{named}{red}
\newcommanddisw{\todo}{todocol}{ToDo: #1}%
\definecolor{migucol}{named}{purple}%
\newcommanddisw{\migucom}{migucol}{{@}comment: #1}%
\newcommanddisw{\miguhigh}{migucol}{#1}%
\usepackage{siunitx}
\usepackage{xspace}
\definecolor{chhucol}{named}{blue}%
\newcommanddisw{\chhucom}{chhucol}{{@}comment: #1}%
\newcommanddisw{\chhuhigh}{chhucol}{#1}%

\definecolor{perocol}{named}{OliveGreen}%
\newcommanddisw{\perocom}{perocol}{{@}comment: #1}%
\newcommanddisw{\perohigh}{perocol}{#1}%

\definecolor{ruhacol}{named}{orange}%
\newcommanddisw{\ruhacom}{ruhacol}{{@}comment: #1}%
\newcommanddisw{\ruhahigh}{ruhacol}{#1}%

	\newcommand{\TempDisplayPreparation}{\disablewr{%
		\section{Draft-State: Comment Color Code}\noindent%
		\todo{Comments: ToDos}\nl%
		\migucom{Comments: Michael Gundall}\nl%

	}}%
\begin{document}%
%Design/Introduction of a Docker-based Virtual Process Controller that enables dynamical reprogramming/reconfiguration%
\title{%
%Using  
Extending Reference Broadcast Infrastructure Synchronization Protocol in IEEE 802.11 as Enabler for the Industrial Internet of Things%Mobile Use Cases
\thanks{This research was supported by the German Federal Ministry for Economic Affairs and Energy (BMWi) within the project FabOS under grant number 01MK20010C. The responsibility for this publication lies with the authors. This is a preprint of a work accepted but not yet published at the IEEE 4th International Conference on Industrial Cyber-Physical Systems (ICPS). Please cite as: M. Gundall, C.Huber, S. Melnyk, and H.D. Schotten: “Extending Reference Broadcast Infrastructure Synchronization Protocol in IEEE 802.11 as Enabler for the Industrial Internet of Things”. In: 2021 4th International Conference on Industrial Cyber-Physical Systems (ICPS), IEEE, 2021.}
}
%
%\input{./organization/IEEE_authors.tex}%
\author{%
\IEEEauthorblockN{%
    Michael Gundall\IEEEauthorrefmark{1}, %
    Christopher Huber\IEEEauthorrefmark{1}, %
    Sergiy Melnyk\IEEEauthorrefmark{1}, %
    and Hans D. Schotten\IEEEauthorrefmark{1}\IEEEauthorrefmark{2} %
    \\%
  %  FirstName1 Lastname1\IEEEauthorrefmark{3} and %
   % FirstName2 Lastname2\IEEEauthorrefmark{4}%
}%
\IEEEauthorblockA{%
    \IEEEauthorrefmark{1}German Research Center for Artificial Intelligence GmbH (DFKI), Kaiserslautern, Germany \\% 
    \IEEEauthorrefmark{2}Department of Electrical and Computer Engineering, Technische Universität Kaiserslautern, Kaiserslautern, Germany %
%     Trippstadter Str. 122\\%
%     67663 Kaiserslautern\\%
	\\%
  %  \IEEEauthorrefmark{3}Institute1, %
  %  Some Subtitle 1 %
%     Optional Address, Germany%
 %   \\%
	%\IEEEauthorrefmark{4}Corporation2, %
  %  Some Subtitle2, %
  %  Some more Subt2 %
%     Optional Address
  %  \\%
    Email: %
        \{michael.gundall, christopher.huber, sergiy.melnyk, hans\_dieter.schotten%
        \}@dfki.de%, %
        \\%
}%
}%


%Telephone: (800) 555--1212, Fax: (888) 555--1212}

\maketitle
\begin{abstract}%
Realizing the \acrlong{iiot2}, more and more mobile use cases will emerge in the industrial landscape, requiring both novel concepts and smooth integration into legacy deployments. %Furthermore, these use cases can be divided into optionally mobile and mandatory mobile, the former considering the use of wireless communication due to soft criteria such as cost savings, and the latter meaning use cases that cannot be covered by wired technologies due to the mobility of the devices. 

Since accurate time synchronization is particularly challenging for wireless devices, we propose a concept for simple but accurate synchronization in IEEE 802.11 \acrlong{wlan} that extends the \acrlong{rbis} protocol, and a suitable integration of IEEE 802.1AS that is part of the \acrlong{tsn} standards. In addition, the concept is evaluated with a testbed using \acrlong{cots} hardware and a realistic discrete automation demonstrator equipped mostly with industrial components. By using the aforementioned devices for wireless communications, this concept can be directly applied in existing industrial solutions, thus achieving the proposed results. It is shown that the achieved synchronicity is suitable for a wide range of mandatory mobile use cases, which are most important for a fully functional \acrlong{iiot2}.
\end{abstract}%
% \fi
\IEEEpeerreviewmaketitle
%
%
%
%
%
% Examples: floats, figure, subfigure (two column float), tables
% -------------------------------------------------------------------
% \input{./organization/backing/IEEETemplateProvidedExamples.tex}
% -------------------------------------------------------------------
% Note that the IEEE does not put floats in the very first column
% - or typically anywhere on the first page for that matter. Also,
% in-text middle ("here") positioning is typically not used, but it
% is allowed and encouraged for Computer Society conferences (but
% not Computer Society journals). Most IEEE journals/conferences use
% top floats exclusively. 
% Note that, LaTeX2e, unlike IEEE journals/conferences, places
% footnotes above bottom floats. This can be corrected via the
% \fnbelowfloat command of the stfloats package.
%
%
%“ ”
%
%#################################################################
%##=========================================================######
%##---------------------------------------------------------######
%#################################################################
%##=========================================================######
%##---------------------------------------------------------######
\section{Introduction}%
\label{sec:Introduction}
%##=========================================================######
%#################################################################
The \gls{iiot2} describes the digitalization of all kinds of production assets that are called \glspl{icps}. With their help, numerous novel use cases that are key enabler for a smart manufacturing, can be realized \cite{8502649}. Furthermore, these use cases can be divided into optionally mobile and mandatory mobile, the former considering the use of wireless communication due to soft criteria such as cost savings, and the latter meaning use cases that cannot be covered by wired technologies due to the mobility of the devices. These use cases enhance the traditional applications in order to ensure the required flexibility of a smart manufacturing. One of the major differences is the requirement of wireless communications in order to allow the increasing number of mobile use cases. Table \ref{tab:Target use cases and selected requirement} sums up and classifies important use cases into classes as well as selected requirements, where we identified \gls{e2e} latency and synchronicity as particularly relevant. In addition, we listed the associated real-time classes for each use case class. Depending on the reference, these are listed from 1-3 or A-C. In the following, we use the latter.

\begin{table}[htbp]
\caption{Novel use cases and selected requirements \cite{8502649,7782431,8731776}}
\begin{center}
\begin{tabular*}{\columnwidth}{p{0.26\columnwidth}ccc}
\cline{1-3} %\hline 
%\begin{tabular}{cccc}
\hline \hline
Use case class & \multicolumn{2}{c}{Requirements} & Real-time \\
\cline{2-3}
 & E2E latency & Synchronicity & class \\
\hline
(I) &  &  &  \\
Remote control, monitoring & 10-100 ms & $\leq$ 1 s & 1 / A \\
\hline
(II) &  &  &  \\
Mobile robotics, process control & 1-10 ms & $\leq$ 1 ms & 2 / B \\
\hline
(III) &  &  &  \\
Closed loop motion control & $<$ 1 ms & $\leq$ 1 $\mu$s & 3 / C \\
\hline \hline
%\cline{1-3} %\hline 
\end{tabular*}
\label{tab:Target use cases and selected requirement}
\end{center}
\end{table}

Use cases that belong to use case class I, such as predictive maintenance and \gls{ar}, belong to the lowest real-time class A and consequently have the lowest latency and synchronization requirements. Here, the time synchronization has to be better than 1\,s. This means that a good time synchronization is not required for these use cases. Since \gls{e2e} latencies of 10\,--\,100\,ms are sufficient, the challenges for these use cases are mostly in the data rates that need to be supported due to the number of sensor nodes or video transmission and coverage of a large area.%, rather than very good time synchronization. 

Use cases belonging to the second class, which include mobile use cases, are the main target of the work in this paper. This class is particularly challenging because of the need for higher performance of time synchronization and \gls{e2e} latency of wireless communications. Particularly demanding are collaborative use cases, where two and more mobile robots have to interact with each other. Here, a synchronicity of ~$<$1~ms is targeted. % challenging are those use cases where multiple mobile devices have a collaborative task, as the most accurate time and state synchronization is required, with better synchronicity leading to faster robot interaction and thus higher productivity. Typically, these use cases require a synchronicity of~$<$1~ms. 

Use case class III has the highest requirements for \gls{e2e} latency and synchronicity and cannot be addressed by current wireless communication technologies. Here both, concepts for combining \gls{tsn}  and \gls{5g} \cite{gundall2021introduction} as well as \gls{tsn} and \gls{wlan} \cite{mildner2019time,adame2019time} are discussed. Since the argument for an introduction of wireless communications in this use case class is mainly the potential of cost savings \cite{7782431,mildner2019time,7883994}, we call these use cases optionally mobile. Thus, we will not address this use case class in our work. % The main arguments for replacing cables with wireless communication in this class are high cost savings \cite{7782431,mildner2019time,7883994}. Therefore, we call these use cases optionally mobile. Since the goal of our research is to leverage existing hardware to enable \gls{iiot2} where wireless communication is mandatory, we will not address use case class III in our work.

Therefore, the following contributions can be found in this paper:
\textbf{
\begin{itemize}
   \item Concept for extending the \gls{rbis} protocol in IEEE 802.11 in order to fulfill the synchronicity required by  \glspl{icps} and the \gls{iiot2}.
  \item Performance evaluation of the method based on a testbed and a discrete automation demonstrator. 
\end{itemize}
}
\newcommand{\wlan}[1][]{IEEE\,802.11#1\xspace}

Accordingly, the paper is structured as follows:  Sec. \ref{sec:Related Work} gives insights into related work on this topic, while an overview on \gls{rbis} protocol and \wlan is given in Sec. \ref{sec:Background}. Furthermore, details on the concept for the integration of \wlan with IEEE 802.1AS using \gls{rbis} protocol are proposed in Sec. \ref{sec:Concept}. This is followed by a performance evaluation based on a testbed consisting mainly of \gls{cots} communication hardware, and a realistic discrete automation demonstrator (Sec. \ref{sec:Testbed}). Finally, the paper is concluded in Sec. \ref{sec:Conclusion}.

%#################################################################
%##=========================================================######
%##---------------------------------------------------------###### 
\section{Related Work}%
\label{sec:Related Work}
%##=========================================================######
%#################################################################
Accurate time synchronization of wireless systems differs significantly from wired systems \cite{fischer2021modular}. This can be caused by changes of the propagation path of the communication signal during operation due to movement or changed channel conditions. Therefore, the time for message propagation between devices is not constant, but can vary greatly, in contrast to cables. Therefore, \cite{9145977,7782431} summarize concepts for time synchronization of wireless systems, the latter focusing on \wlan \gls{wlan}. In addition, a concept of an improved time synchronization by estimating the distance to the \gls{ap} using the station's receive power \cite{9145977} was proposed. Furthermore, a concept for integrating \gls{ptp} and \wlan using the \gls{rbis} protocol was presented \cite{6489696} and evaluated \cite{7018946}. This method takes advantage of the broadcast nature of the wireless medium and is well suited for simple but accurate time synchronization of wireless devices. Therefore, this method was also studied for synchronization of \gls{3gpp} 4G and \gls{5g} systems \cite{gundall2020integration}. Since all approaches built on the \gls{rbis} protocol include the assumption that all \glspl{icps} are connected to the same \gls{ap} or base station, this work extends these concepts by this feature.

%#################################################################
%##=========================================================######
%##---------------------------------------------------------###### 
\section{Background}%
\label{sec:Background}
%##=========================================================######
%#################################################################
%To meet the stringent demands of upcoming use cases, several novel technologies are being introduced. This section provides an overview of the most relevant technologies that have been used for our work.

\subsection{RBIS Protocol}
\begin{figure}[htbp]
\centerline{\includegraphics[width=.83\columnwidth]{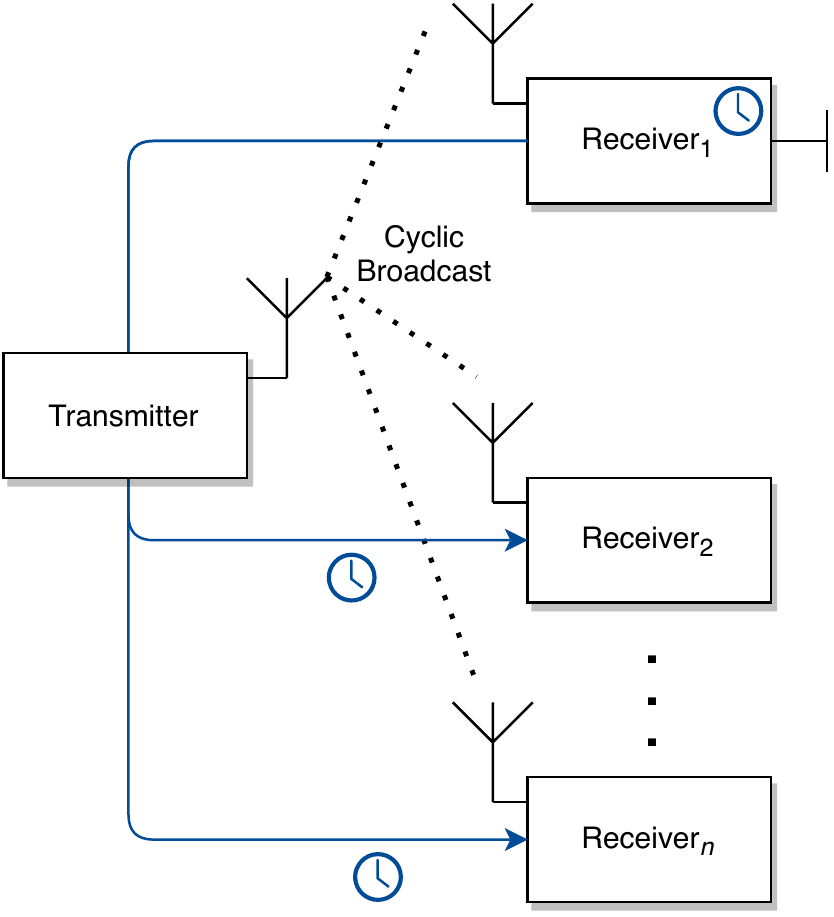}}
\caption{System components for applying RBIS protocol.}
\label{fig:RBIS protocol}
\end{figure}

RBIS is a master/slave clock synchronization protocol that was developed for wireless communication systems using infrastructure-based communication \cite{4579760} and is shown in Figure \ref{fig:RBIS protocol}. 
This means, that the system has a fixed transmitter, such as an \gls{ap} or a \gls{bs}, that cyclically broadcasts messages to all receivers that should be synchronized. Next, each receiver makes a timestamp for each incoming message, and thus can calculate the offset to the master that informs each subscribed slave of the correct time compared to each time point, in this case Receiver\textsubscript{1}. This mechanism is also called receiver/receiver paradigm. With Receiver\textsubscript{1} connected to the wired backbone of a facility, it can act as a time slave for time synchronization protocols such as specified in \gls{tsn} and forward that time to all receivers connected to the same base station using \gls{rbis} protocol.

\subsection{IEEE 802.11}

\wlan defines a family of standards for \gls{wlan}, such as \wlan[a/b/g/n/ac/ad]. In the most recent one, which is also called \mbox{Wi-Fi}~5, two frequency bands can used. While on the data rate on the \SI{2.4}{\GHz} frequency band is limited to \SI{600}{Mbps},  a theoretical rate of \SI{6.933}{Gbps} can be reached using the \SI{5}{\GHz} band in combination with $8\times8$-\gls{mimo} \cite{melnyk:2017}. In order to further increase the data rate, in a newer version of the \wlan standard, which is called \mbox{Wi-Fi}~6, also  physical layer technologies are adopted.

Additionally, in \wlan \gls{wlan} two different operating modes are defined. These modes are called infrastructure and ad-hoc mode. Since the devices can only communicate peer-to-peer, this mode is not used in most installations. Therefore, we only consider infrastructure mode, where stations communicate with each other using an \gls{ap}, in our investigations. Thus, our concept can only applied to this mode.

\section{Extended RBIS protocol}%
\label{sec:Concept}
%##=========================================================######
%#################################################################
For the application of \gls{rbis} protocol in infrastructure mode based \gls{wlan} deployments, a broadcast message is required, that is transmitted by an \gls{ap} to each station. Furthermore, it has to contain an unique and non repeatable identifier, in order to be able to differentiate these messages. This is important, since each station produces a timestamp for each incoming broadcast messages that is required to calculate the time offset between stations. 

%As already mentioned, most \gls{wlan} deployments use the infrastructure mode, in which the stations do not communicate with each other directly, but via an \gls{ap}. To use the \gls{rbis} protocol for time synchronization of the stations, the following conditions must be satisfied:  
%\begin{itemize}
   % \item The message should arrive at each station at the same time
   % \item Each message should have a unique identifier that is not repeatable
    %\item The frequency of message transmission should be high
%\end{itemize}

%In addition to the user data that applications send from one station to another via an \gls{ap}, there are also control and management messages that are transmitted by the \gls{ap}, for example to share metadata. One of the management messages sent by each \gls{ap} is shown in Figure \ref{fig:Beacons}. 

\begin{figure}[tb]
%\centerline{\includegraphics[width=\columnwidth]{figures/Beacon Frame PCAP.png}}
\centerline{\includegraphics[width=\columnwidth]{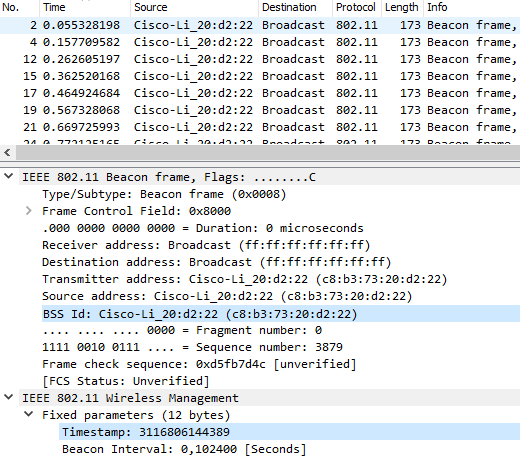}}
\caption{Beacon frames captured by a packet analyzer.}
\label{fig:Beacons}
\end{figure}

Therefore, we use the so called beacon frames that are part of the management frames and include several data values, whereby an exemplary beacon frame is shown in Figure \ref{fig:Beacons}. 
Each beacon frame contains besides the the SSID of the \gls{ap}, the interval of the periodic beacon frame transmission. This value is called \gls{bi} and is by default 100, i.\,e. 102.4 ms, whereby a lower value represents a higher broadcast frequency. How often the beacon frame should be transmitted is a trade off between synchronization accuracy and required data rate, since latter will be reduced with an higher amount of beacon frames transmitted. In addition, each beacon frame includes the beacon frame timestamp. This value states the time of the \gls{ap}, which is typically the time since the \gls{ap} was turned on.

\begin{figure*}[tb]
\centerline{\includegraphics[width=.7\textwidth]{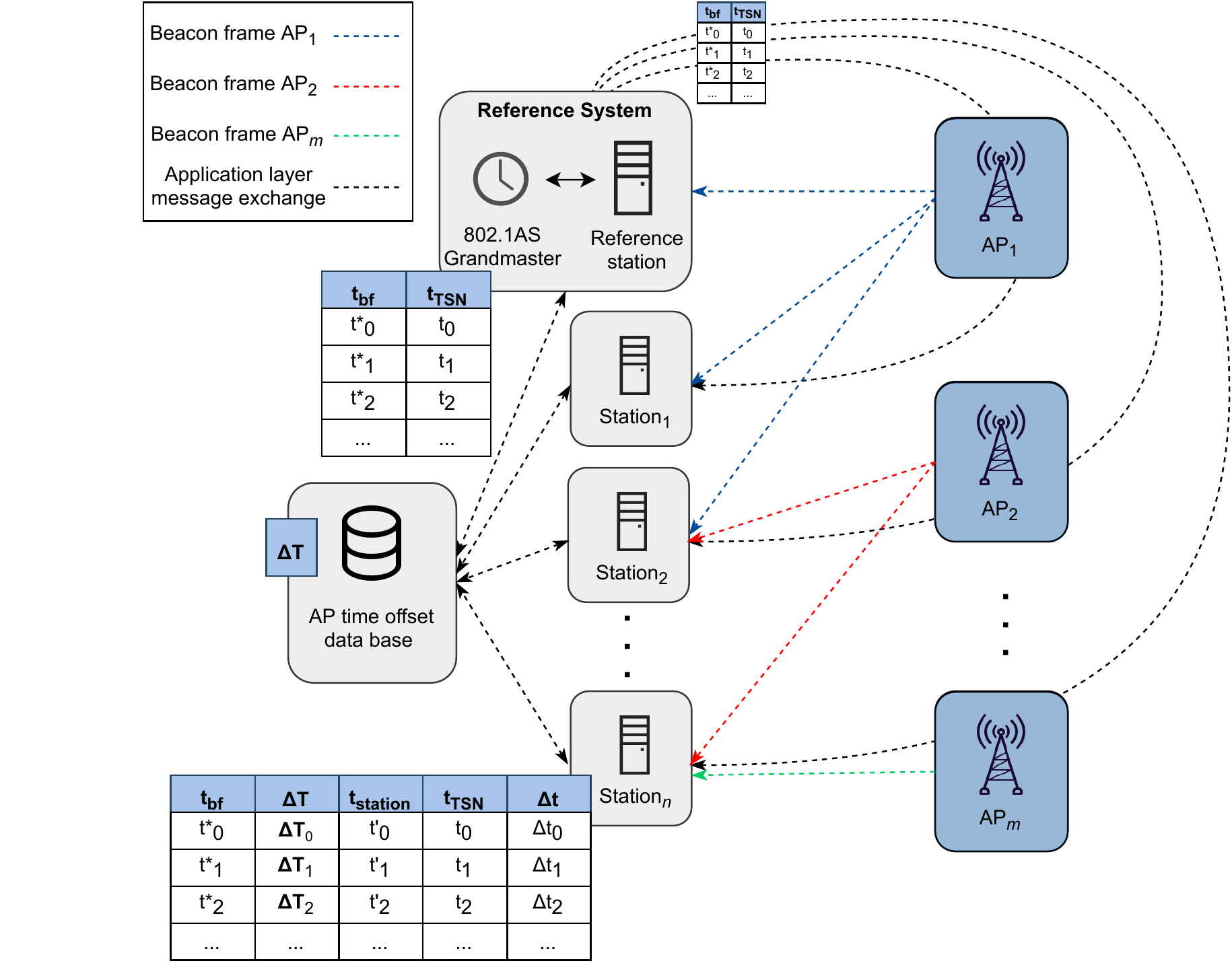}}
\caption{Concept for the distribution of the \gls{tsn} time in the \gls{wlan} network using extended \gls{rbis} protocol.}
\label{fig:concept}
\end{figure*}

%These messages are called beacon frames and contain the SSID of the \gls{ap}, the time interval of the transmission, and the timestamp of the beacon, i.\,e. the time that elapsed since the \gls{ap} was powered. By default, the \gls{bi} is 100, i.\,e. 102.4 ms. If the synchronization should be improved, this value can easily be adopted, but with a higher number of management frames transmitted, the maximum data rate will be reduced. In a realistic industry landscape, not only one but multiple \glspl{ap} are in range of each station in order to guarantee seamless coverage. Thus, a station usually receives beacon frames from multiple \glspl{ap}. To separate them, it is useful to filter by the BSS ID, which corresponds to the MAC address of the \gls{ap} and is also transmitted in the beacon frame. Because of the characteristics mentioned above, beacon frames are thus well suited for the \gls{rbis} protocol. 
As mentioned earlier, the \gls{rbis} protocol implies the assumption that all \glspl{icps} are connected to the same \gls{ap}, which is not suitable for covering the entire factory floor. To enable scenarios where \glspl{icps} are connected to different \glspl{ap}, our concept for extending \gls{rbis}, which is shown in Figure \ref{fig:concept}, also integrates this feature.  
%0.67= \footnotesize, 0.83 = normal
The concept consists of multiple \glspl{ap} and stations, whereby one of them is specified as “Reference Station". This special station is part of the “Reference System" and is connected to the wired \gls{tsn} network and can therefore not be mobile. The reason for this is that the station is time synchronized by the \gls{gm}, which is located in the wireline \gls{tsn} network. Therefore, the station that is used as Reference Station has to support time synchronization based on IEEE 802.1AS.

In the first step, the Reference System timestamps each incoming beacon timestamp with its \gls{tsn} time. Afterwards, it sends the correct timestamp for each beacon frame to all other stations that shlould be time synchronized.

Next, each station couples the timestamp of its local time for each incoming beacon frame with the beacon frame timestamp. These tuples can then be used, in order to calculate the time offset. 

After this this procedure, each station connected to the same \gls{ap} is time synchronized, e.\,g. Station\textsubscript{1} and Station\textsubscript{2} in Figure \ref{fig:concept}. If  Station\textsubscript{2} would move to an other location on the factory floor, AP\textsubscript{1} might be not longer be in its range and might be connected to an other \gls{ap}, e.\,g., AP\textsubscript{2}. If we assume that Station\textsubscript{2} is in range of both \glspl{ap}, and publishes their time offsets to a database, Station\textsubscript{n} can be time synchronized with Station\textsubscript{1} even if they are not connected to the same \gls{ap}. Next, the equation for the correction of the clock for each sation can be determined. 
 
%\begin{align*}
\begin{equation}
\label{eq:sn}
%t_{TSN} = t_{ TSN }[SN] - t_{ Station }[SN] + t_{ Station }[current] 
t_\text{TSN} = t_\text{TSN}[t_\text{bf}] - t_\text{station}[t_\text{bf}] + t_\text{ station }[t] + \Delta t_\text{AP}
%\end{align*}    
\end{equation}

Here,  $t_\text{bf}$ is the beacon frame timestamp, %$t$\textsubscript{TSN}$[sn]$  %$t$\textsubscript{TSN}$[sn]$ 
% $t_{TSN}[SN]$ 
$t_\text{TSN}[t_\text{bf}]$
is the time of the Reference Station for at the arrival of the beacon frame, %$t$\textsubscript{UE}$[sn]$ 
% $t_{UE}[SN]$ 
$t_\text{station}[t_\text{bf}]$
is the local time of the station that should be synchronized for the same beacon frame,  %$t$\textsubscript{UE}$[current]$ 
%$t_{ UE }[current]$ 
$t_\text{station}[t]$ 
 is the current time of the station, and
 $\Delta t_\text{AP}$ is the offset of both \glspl{ap}. If there is a lot of mobility in the system, it may be advantageous for the AP time offset database to publish the offsets between each of the \glspl{ap} cyclically, rather than being requested acyclically by each station. An offset matrix $\Delta T$ with dimensions $m \times m$, $i$ rows, and $j$ columns can be formed, where $m$ is the number of \glspl{ap}: %, an offset matrix , as shown in Equation \ref{eq:T}. 
\begin{equation}
\label{eq:T}
\Delta T = \left(\begin{IEEEeqnarraybox*}[][c]{,c/c/c/c,}
\Delta t_{11} & \Delta t_{12} & \cdots & \Delta t_{1m}\\
\Delta t_{21} & \Delta t_{22} & \cdots & \Delta t_{2m}\\
 \vdots &  \vdots & \vdots & \vdots \\
\Delta t_{m1} & \Delta t_{m2} & \cdots & \Delta t_{ij}%
\end{IEEEeqnarraybox*}\right)
\end{equation}
Since each $\Delta t_{ij}=\Delta t_{ji}$ and each $\Delta t_{ii}=0$, the number of entries in the database $n_{entries}$ and thus the payload of the data packets can be calculated by the upper triangular matrix of a square matrix minus the values in its main diagonal: %(see Equation \ref{eq:number}).
\begin{equation}
\label{eq:number}
n_{entries} = \frac{m(m+1)}{2}-m= \frac{m(m-1)}{2}
\end{equation}
Depending on the number of handovers of stations between \glspl{ap} and the number of \glspl{ap}, the calculated $n_{entries}$ can be used to apply a tailored strategy, e.\,g., higher accuracy due to cyclic publishing or less consumed bandwidth when the channel is busy.

%\migucom{Data base has to be updated with interval dependant on clock drift of APs, maybe FOrmula with clock drift  and overall accuracy should be written here}

%\migucom{Formel für Beacon Interval und Follow up Interval basierend auf Clock-Drift von 1 oder mehreren APs und Stations}

%\migucom{Stolen: Since the area covered by real-world BSSs is not very large—the maximum distance allowed between the STAs and the AP is in the order of one hundred meters at most—the rror incurred because propagation delays are not compensated below 1 s) is typically lower than the jitter caused by s/w imestamps. Therefore, it could be neglected in s/w-based mplementations. Using h/w timestamps in RBIS decreases oth in-node latencies and the related jitters noticeably, but the rror due to the uncompensated propagation delay remains. lthough the inability of RBIS to estimate propagation elays is a clear limitation, it can be overcome in several ractical cases.  ore APs than evcery 100m, vllt auf eployment Konzept von Oscar oder so.}

%#################################################################
%##=========================================================######
%##---------------------------------------------------------###### 
\section{Testbed \& Evaluation}%
\begin{figure}[tb]
%\centerline{\includegraphics[width=\columnwidth]{figures/Testbed_WiFi.pdf}}
\centerline{\includegraphics[width=\columnwidth]{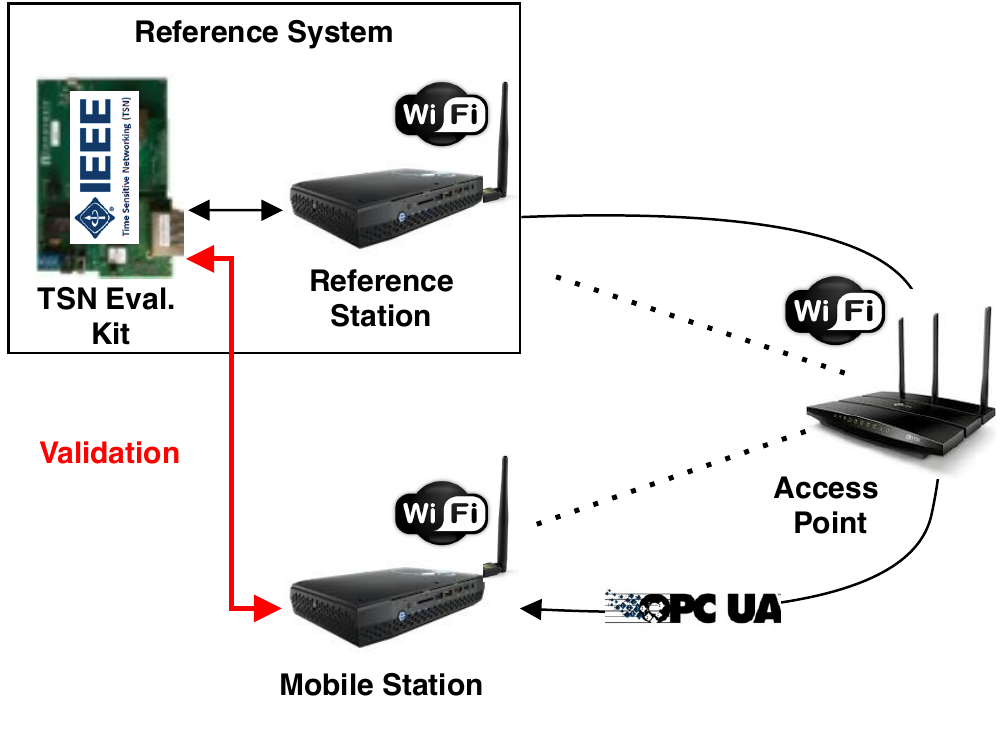}}
\caption{Testbed for measuring the accuracy of RBIS protocol in \mbox{Wi-Fi}.}
\label{fig:Testbed}
\end{figure}

\label{sec:Testbed}
%##=========================================================######
%#################################################################
In order to validate our approach, it is evaluated on the basis of the testbed, which is shown in Figure \ref{fig:Testbed}. Furthermore, the testbed is equipped with the \gls{cots} components specified in Table~\ref{tab:hardware}. This guarantees the possibility to integrate the concept in existing \gls{wlan} installations without any hardware change. 

%This section aims to evaluate the proposed concept. Therefore, Figure \ref{fig:Testbed} shows the testbed, the evaluation was performed with. It mainly consists of the \gls{cots} components, listed in Table~\ref{tab:hardware}.
%
%0.67= \footnotesize, 0.83 = normal
\begin{table}[tb]
\caption{Hardware configurations}
\begin{center}
\begin{tabular*}{\columnwidth}{|p{0.26\columnwidth}|c|p{0.5\columnwidth}|}
\cline{1-3} %\hline 
\textbf{\textit{Equipment}} & \textbf{\textit{QTY}} & \textbf{\textit{Specification}}\\
\cline{1-3} %\hline 
Mini PC & 2 & Intel Core i7-8809G, 2x16 GB DDR4, Intel i210-AT \& i219-LM Gibgabit NICs, Ubuntu 18.04.3 LTS 64-bit, \linebreak Linux 4.18.0-18-lowlatency  \\
%& & Processor: Intel Core i7-8809G  \\
%& & Kernel: Linux 4.19.103-rt42  \\
\cline{1-3} %\hline 
\mbox{Wi-Fi} Adapter & 2 & USB, \mbox{Wi-Fi} 5  \\
\cline{1-3} 
\mbox{Wi-Fi} Router & 1 & IEEE~802.11ac, \mbox{Wi-Fi} 5 \\
\cline{1-3}%\hline 
TSN Evaluation Kit & 1 & RAPID-TSNEK-V0001, IEEE~802.1AS \\%, 802.1Qbv, 802.1Qci, 802.1CB, 802.1Qcc, 802.1Qbu / 802.3br \\
\cline{1-3} %\hline 
\end{tabular*}
\label{tab:hardware}
\end{center}
\end{table}

In the testbed, two \glspl{pc} are connected via an \mbox{Wi-Fi} router. Since \mbox{Wi-Fi} modules are typically multiplexing in order to listen to \mbox{Wi-Fi} signals in various channels, not all beacon frames are received by default. Therefore, the \mbox{Wi-Fi} network interfaces have to be set to a special monitoring mode. Using aricrack-ng module \cite{aircrack}, a complete channel can continuously monitored. However, this leads to the drawback that it is not possible anymore to send IPv4 messages over the network. Since this is required to submit offset correction messages, an additional \mbox{Wi-Fi} adapter per \gls{pc} was added. 

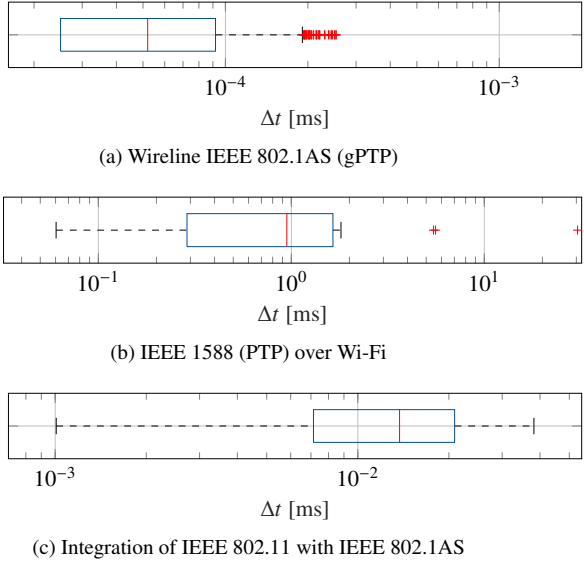
\begin{figure}[tb]
	\centering
		\subfloat[Wireline IEEE 802.1AS (gPTP)]{\resizebox{\columnwidth}{!}{%
\definecolor{mycolor1}{rgb}{0.00000,0.3,0.6}
\begin{tikzpicture}

\begin{axis}[%
width=3.5in,
height=0.4in,
scale only axis,
unbounded coords=jump,
xmode=log,
xmin=-0.000001,
xmax=0.002,
xlabel style={font=\color{white!15!black}},
xlabel={$\Delta t$ [ms]},
ymin=0.7,
ymax=1.3,
ytick={1},
yticklabels={\empty},
ylabel style={font=\color{white!15!black}},
ylabel={~},
axis background/.style={fill=white},
xmajorgrids,
ymajorgrids,
]
\addplot [color=black, dashed, forget plot]
  table[row sep=crcr]{%
0.000092	1\\
0.000191	1\\
};
\addplot [color=black, dashed, forget plot]
  table[row sep=crcr]{%
0	1\\
0.000025	1\\
};
\addplot [color=black, forget plot]
  table[row sep=crcr]{%
0.000191	0.925\\
0.000191	1.075\\
};
\addplot [color=black, forget plot]
  table[row sep=crcr]{%
0	0.925\\
0	1.075\\
};
\addplot [color=mycolor1, forget plot]
  table[row sep=crcr]{%
0.000025	0.85\\
0.000092	0.85\\
0.000092	1.15\\
0.000025	1.15\\
0.000025	0.85\\
};
\addplot [color=red, forget plot]
  table[row sep=crcr]{%
0.000052	0.85\\
0.000052	1.15\\
};
\addplot [color=black, draw=none, mark=+, mark options={solid, red}, forget plot]
  table[row sep=crcr]{%
0.000193	1\\
0.000194	1\\
0.000194	1\\
0.000194	1\\
0.000194	1\\
0.000195	1\\
0.000196	1\\
0.000197	1\\
0.000197	1\\
0.000198	1\\
0.000199	1\\
0.000201	1\\
0.000202	1\\
0.000202	1\\
0.000202	1\\
0.000203	1\\
0.000204	1\\
0.000206	1\\
0.000206	1\\
0.000209	1\\
0.000210	1\\
0.000214	1\\
0.000215	1\\
0.000215	1\\
0.000216	1\\
0.000219	1\\
0.000220	1\\
0.000221	1\\
0.000221	1\\
0.000230	1\\
0.000231	1\\
0.000238	1\\
0.000239	1\\
0.000243	1\\
0.000244	1\\
0.000246	1\\
0.000250	1\\
0.000251	1\\
0.000253	1\\
0.000254	1\\
};
\end{axis}

\end{tikzpicture}%
}\label{fig:Sub_TSN}}

	\centering
		\subfloat[IEEE 1588 (PTP) over \mbox{Wi-Fi}]{\resizebox{\columnwidth}{!}{%
\definecolor{mycolor1}{rgb}{0.00000,0.3,0.6}%
\begin{tikzpicture}

\begin{axis}[%
width=3.5in,
height=0.4in,
scale only axis,
unbounded coords=jump,
xmode=log,
xmin=-1.45863275,
xmax=31.95995575,
xlabel style={font=\color{white!15!black}},
xlabel={$\Delta t$ [ms]},
ymin=1.7,
ymax=2.3,
ytick={1},
yticklabels={\empty},
ylabel style={font=\color{white!15!black}},
ylabel={~},
axis background/.style={fill=white},
xmajorgrids,
ymajorgrids,
]
\addplot [color=black, dashed, forget plot]
  table[row sep=crcr]{%
1.6432615	2\\
1.808659	2\\
};
\addplot [color=black, dashed, forget plot]
  table[row sep=crcr]{%
0.060394	2\\
0.2879375	2\\
};
\addplot [color=black, forget plot]
  table[row sep=crcr]{%
1.808659	1.925\\
1.808659	2.075\\
};
\addplot [color=black, forget plot]
  table[row sep=crcr]{%
0.060394	1.925\\
0.060394	2.075\\
};
\addplot [color=mycolor1, forget plot]
  table[row sep=crcr]{%
0.2879375	1.85\\
1.6432615	1.85\\
1.6432615	2.15\\
0.2879375	2.15\\
0.2879375	1.85\\
};
\addplot [color=red, forget plot]
  table[row sep=crcr]{%
0.9475505	1.85\\
0.9475505	2.15\\
};
\addplot [color=black, draw=none, mark=+, mark options={solid, red}, forget plot]
  table[row sep=crcr]{%
5.459047	2\\
5.591106	2\\
30.440929	2\\
};
\end{axis}
\end{tikzpicture}%
}\label{fig:Sub_PTP}}

		\subfloat[Integration of IEEE 802.11 with IEEE 802.1AS]{\resizebox{\columnwidth}{!}{%
\definecolor{mycolor1}{rgb}{0.00000,0.3,0.6}
\begin{tikzpicture}

\begin{axis}[%
width=3.5in,
height=0.4in,
scale only axis,
unbounded coords=jump,
xmode=log,
xlabel style={font=\color{white!15!black}},
xlabel={$\Delta t$ [ms]},
ymin=0.7,
ymax=1.3,
ytick={1},
yticklabels={\empty},
ylabel style={font=\color{white!15!black}},
ylabel={~},
axis background/.style={fill=white},
xmajorgrids,
ymajorgrids,
]
\addplot [color=black, dashed, forget plot]
  table[row sep=crcr]{%
%0.119715	1\\
%0.19981	1\\
0.0208905   1\\
0.038196    1\\
};
\addplot [color=black, dashed, forget plot]
  table[row sep=crcr]{%
0.001008	1\\
%0.0332	1\\
0.00713775  1\\
};
\addplot [color=black, forget plot]
  table[row sep=crcr]{%
%0.19981	0.925\\
%0.19981	1.075\\
0.038196 	0.925\\
0.038196 	1.075\\
};
\addplot [color=black, forget plot]
  table[row sep=crcr]{%
0.001008	0.925\\
0.001008	1.075\\
};
\addplot [color=mycolor1, forget plot]
  table[row sep=crcr]{%
%0.0332	0.85\\
%0.119715	0.85\\
%0.119715	1.15\\
%0.0332	1.15\\
%0.0332	0.85\\
0.00713775	0.85\\
0.0208905	0.85\\
0.0208905	1.15\\
0.00713775	1.15\\
0.00713775	0.85\\
};
\addplot [color=red, forget plot]
  table[row sep=crcr]{%
0.0137405	0.85\\
0.0137405	1.15\\
};
\end{axis}
\end{tikzpicture}%
}\label{fig:Sub_SFN}}

\caption{Readings of the measurements for the evaluation of our concept compared to wireline and state of the art solutions.}
\label{fig:results}
\end{figure}

%The system is composed of two mini \glspl{pc} wirelessly connected to a \mbox{Wi-Fi} router that served as \gls{ap}. To receive beacon frames, the \gls{wlan} network interfaces must be set to "monitor mode".  This can be achieved with the help of the aricrack-ng module \cite{aircrack}.  After this, the entire physical channel of the interface can be monitored, however IPv4 connectivity is lost. To be able to publish the time offset from the Reference Station using OPC UA multicast, an additional \mbox{Wi-Fi} adapter was connected via USB. 
Furthermore, a \gls{tsn} evaluation kit is used. It is connected to the reference station and serves as \gls{tsn} \gls{gm}. In order to allow time synchronization, the reference station uses Linux~PTP module \cite{cochran2015linux}. Linux~PTP is an open source implementation of IEEE 802.1AS and 802.1AS-REV specifications. In addition, the \gls{tsn} \gls{gm} is connected to the mobile station. With the help of the \gls{gm} the clock of the mobile station can be compared to the correct \gls{tsn} time and both, clock offset and accuracy of the synchronization can be determined.
%Furthermore, the mini \gls{pc}, which serves as a reference station, is connected to the \gls{tsn} Evaluation Kit, which supports the IEEE 802.1AS and 802.1AS-REV specifications.  Since the Reference Station is running the Linux PTP program \cite{cochran2015linux}, which supports the IEEE 802.1AS specification for \gls{tsn} end stations, it can be synchronized by the \gls{tsn} Evaluation Kit. To be able to measure the accuracy of the synchronization, the \gls{tsn} Evaluation Kit is also connected to the second station and synchronizes the hardware clock of one of the built-in network interfaces. In this way, the time difference between the internal clock and the hardware clock can be measured. 
Therefore, Figure \ref{fig:results} shows the results for the measurements of three different scenarios. First, the synchronization precision that can be achieved with this equipment using wireline is shown in Figure \ref{fig:Sub_TSN}. Here, the median is 52~ns and the maximum value 254~ns. It can be seen, that the achieved synchronicity is sufficient to address each of the proposed use case classes.

The accuracy of transmitting IEEE 1588 protocol directly via plain \mbox{Wi-Fi}, as it is done in wireline systems, is shown in Figure \ref{fig:Sub_PTP}. Here, it can be seen, that this method is not suitable, even if at least some use cases of use case class II could be realized if only the median value is considered, which is with $\approx$0.95~ms below the limit of~$<$1~ms. However, there are too many outliers that are located in a range up to several ms. Therefore, the required synchronicity is not fulfilled for industrial use cases of use case class II. Last but not least, Figure \ref{fig:Sub_SFN} shows the values by using \gls{rbis} protocol for synchronization of the stations. Here, the median is  13~µs and the highest measured value is $<$40~µs. Therefore, the synchronicity requirements of use case classes II and III can be fulfilled. Therefore, by using only \gls{cots} equipment, the synchronization requirements for all mandatory mobile use cases can be met. 

\begin{figure}[tb]
\centerline{\includegraphics[width=\columnwidth]{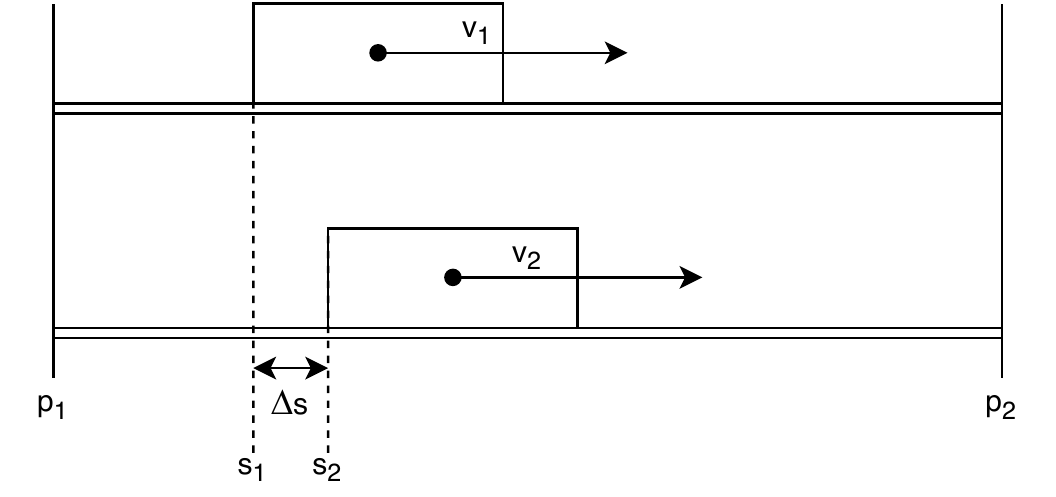}}
\caption{Concept for the identification of the mechanical offset and time synchronization \cite{gundall2020integration}.}
\label{fig:DemoConcept}
\end{figure}

\begin{figure*}[htbp]
\centerline{\includegraphics[scale=1]{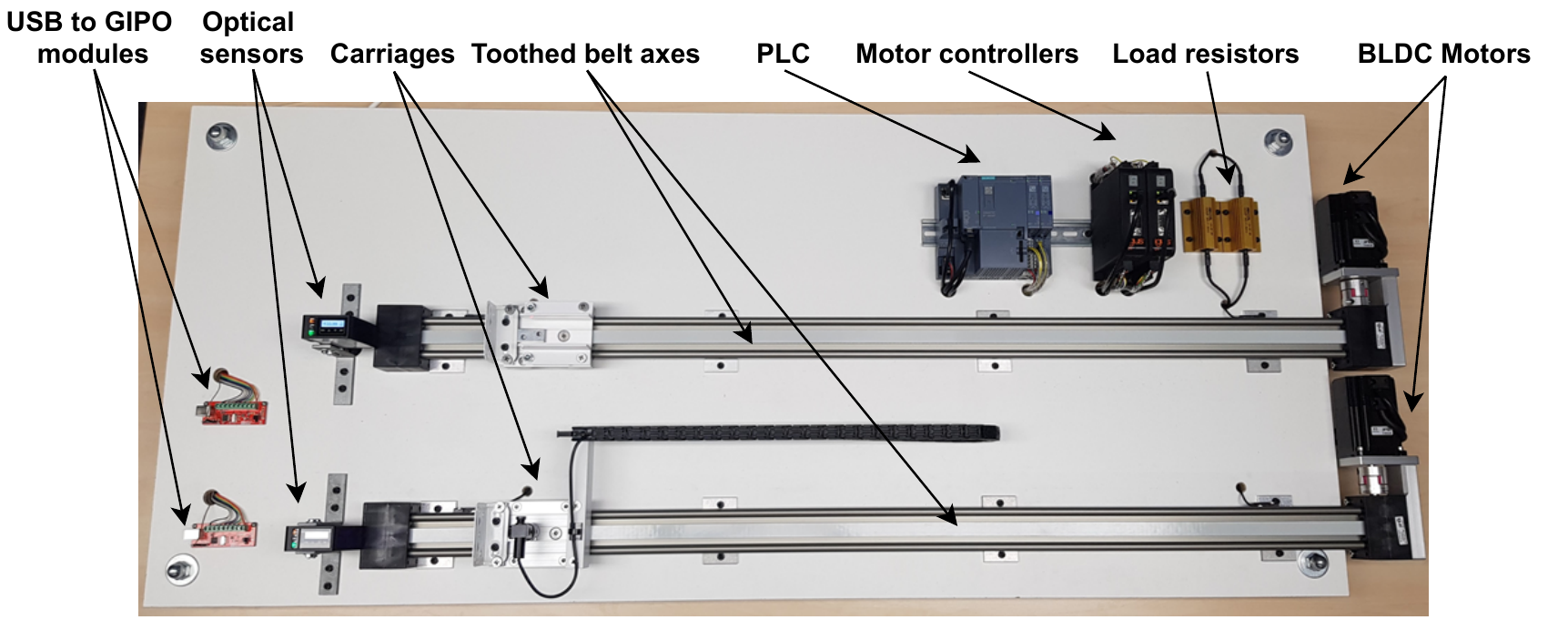}}
%\centerline{\includegraphics[width=\textwidth]{figures/HardwareDemo_Bold.pdf}}
\caption{Setup of the discrete automation demonstrator that was also basis of the investigations in \cite{gundall2020integration}.}
\label{fig:Demo}
\end{figure*}

In order to show the applicability of the proposed concept and its performance, a discrete automation demonstrator is proposed (see Figure \ref{fig:Demo}). It is based of a simultaneous one-dimensional motion of two carriages on linear axes, as it is shown in Figure \ref{fig:DemoConcept}. 
Therefore, each carriage is controlled by a different \mbox{Wi-Fi} station. Thus, if the mechanical movements and conditions are equal, such as the same start (p\textsubscript{1}) and end point (p\textsubscript{2}) as well as the same acceleration and velocity, the mechanical synchronicity can be calculated. Furthermore, this value can be transformed to the time synchronicity, if Equations \ref{eq:position} \& \ref{eq:time} are applied. 

%The idea is that both carriages move together from a common start position $p_1$ to a defined end point $p_2$. If the acceleration as well as the speed of both carriages are the same, the mechanical synchronicity can be determined by measuring the position offset. Hereby, the highest value is achieved, when the maximum speed is reached (see Equation \ref{eq:position} \& Equation \ref{eq:time}). If the speed of every carriage is known, the time delay $\Delta t$ with which the two carriages started their movement can be determined. This corresponds to the time synchronicity.

\begin{equation}
\label{eq:position}
\Delta s(t) = s_1(t) - s_2(t)
\end{equation}

\begin{equation}
\label{eq:time}
\Delta t_{max} = \frac{\Delta s_{max}}{v_{max}} 
\end{equation}

\begin{figure}[tb]
\resizebox{\columnwidth}{!}{%
\input{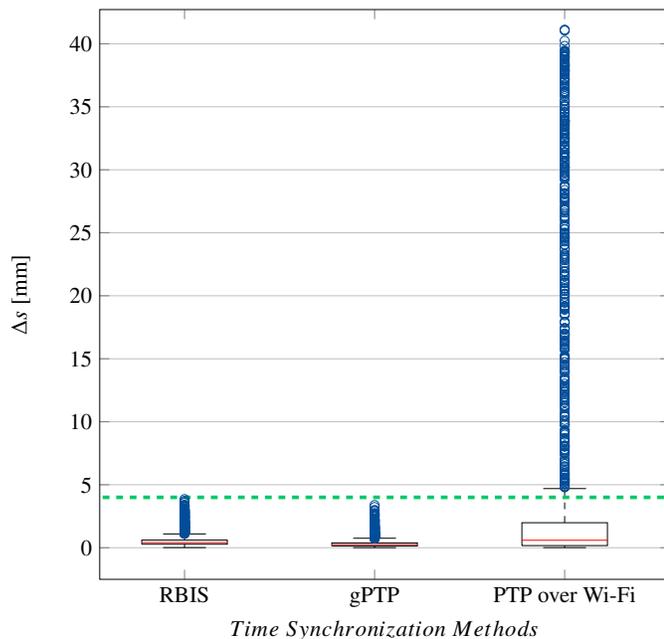}}
%centerline{\includegraphics[width=\columnwidth]{figures/TSN_SFN_PTP.jpg}}
\caption{Readings for the measurements of the the three different synchronization methods.}
\label{fig:One}
\end{figure}

Since the demonstrator shown in Figure \ref{fig:Demo} was first proposed in \cite{gundall2020integration} and both the functionalities and the hardware components are explained in detail in that paper, we only give an overview of the most important components here. It consists of two toothed belt axes, each connected to a \gls{bldc} motor and is moving a carriage, which can reach a maximum speed of $|v_{max}|~=4~\frac{m}{s}$ and an acceleration of $a_{max}~=\pm30~\frac{m}{s^2}$. %The latter ensures that the maximum speed is given over a major part of the 1m stroke. 
Furthermore, the \gls{bldc} motors are controlled by two motor controllers of the same manufacturer. Furthermore, the motor controllers have several GIPOs. With these GPIOs, setpoints for target position can be set and a movement can be triggered. %can be used to specify both setpoints for the target positions and the start of the movement. 
Since we use mini \glspl{pc} as \mbox{Wi-Fi} stations, it is possible to connect them to the motor controllers using USB to GPIO modules. These modules are also depicted in Figure \ref{fig:Demo}.  In order to detect the positions of the carriages, two high-precision sensors were installed. In high-precision mode, in which a Kalman filter uses a time series of several values to reduce the noise, the resolution is about 3-63~$\mu$m depending on the position of the carriages. %This precision is sufficient to validate our concept. 
In order to measure the values of the position sensors, a \gls{plc} is installed, whereby each of the high-speed analog input units has a 12 bit resolution. For the validation, we performed measurements for the same three setups as for the synchronization before: (1) synchronizing two mini \glspl{pc} with the standard IEEE 802.1AS (\gls{gptp}) synchronization mechanism without any wireless communication, (2) transmitting the IEEE 1588 (\gls{ptp}) messages directly via \mbox{Wi-Fi}, and (3) applying the concept, proposed in this paper. The results are shown in Figure \ref{fig:One} and Table \ref{tab:2}. 
\begin{table}[!t]
\caption{Median and maximum values of the measurements for the evaluation of our concept.}
\begin{center}
\begin{tabulary}{\columnwidth}{|p{0.25\columnwidth}|C|C|C|}
\hline 
 & \textbf{RBIS protocol} & \textbf{gPTP} & \textbf{PTP over \mbox{Wi-Fi}} \\
%\cline{2-4} 
  \hline
 \textbf{Median [mm]} & 0.41 & 0.25 & 0.59 \\
  \hline
 \textbf{Max [mm]} & 3.86 & 3.39 & 41.12\\
  \hline
\end{tabulary}
\label{tab:2}
\end{center}
\end{table}
The green horizontal line at 4~mm indicates the limit for use case class II for Equation \ref{eq:time} resolved to $\Delta s_{max}$ for $|v_{max}|~=4~\frac{m}{s}$ and $\Delta t_{max~}=1~ms$. Thus, several things can be observed. First, the accuracy required for the use case class II can be satisfied in a consistent manner. Second, due to the uncertainty of the mechanics of the system, no difference can be observed between wired and wireless transmission when the synchronization method of our system is used. Finally, it is shown that the identical use of wired synchronization mechanisms for wireless systems, such as \mbox{Wi-Fi}, does not result in a deterministic synchronization. In this case, the peak value for synchronization of PTP over \mbox{Wi-Fi} is more than 10 times higher compared to both other methods.

%\end{comment}
%#################################################################
%##=========================================================######
%##---------------------------------------------------------######
\section{Conclusion}%
\label{sec:Conclusion}
%##=========================================================######
%#################################################################
In this work, we presented the requirements and challenges for time synchronization of wireless use cases. For this purpose, we introduced use case classes and divided them into optional mobile and mandatory mobile. Furthermore, we proposed a concept capable of addressing these challenges. Afterwards, an evaluation using a testbed consisting of \gls{cots} components as well as a validation using a discrete automation demonstrator was done. It was shown that by applying the proposed method, accurate time synchronization of all mandatory mobile use cases can be achieved, by using \gls{cots} \mbox{Wi-Fi} components. Thus, a smooth integration into existing installations can be realized. 

%\section*{Acknowledgment}%
%##=========================================================######
%#################################################################
% research was supported by the German Federal Ministry for Economic Affairs and Energy (BMWi) within the project FabOS under grant number 01MK20010C. The responsibility for this publication lies with the authors.



%\nocite{*}
\printbibliography%

%#################################################################
%##=========================================================######
%##---------------------------------------------------------######
% - - - Tempory active while working state. Deactivated before \begin{document}
%##=========================================================######
%#################################################################
\TempDisplayPreparation
\end{document}